\documentclass[10pt,a4paper,twocolumn]{article}
\usepackage[T1]{fontenc}
\usepackage[utf8]{inputenc}
\usepackage{lmodern}
\usepackage[margin=18mm,columnsep=6.5mm]{geometry}
\usepackage{amsmath,amssymb,graphicx,booktabs,array,microtype}
\graphicspath{{./}}
\usepackage[numbers,sort&compress]{natbib}
\usepackage{caption,xcolor,stfloats,placeins,fancyhdr,titlesec,flushend}
\usepackage[colorlinks=true,linkcolor=blue!45!black,citecolor=blue!45!black,urlcolor=blue!45!black]{hyperref}
\titleformat*{\section}{\large\bfseries\raggedright}
\titleformat*{\subsection}{\normalsize\bfseries\raggedright}
\titlespacing*{\section}{0pt}{2.5ex plus 1ex minus 0.2ex}{1.25ex plus 0.2ex}
\titlespacing*{\subsection}{0pt}{2.0ex plus 0.7ex minus 0.2ex}{0.9ex plus 0.2ex}
\newcommand{\jpsi}{J/\psi}

\newcommand{\GeV}{\,\mathrm{GeV}}
\newcommand{\CS}{\,{}^3S_1^{[1]}}
\newcommand{\COs}{\,{}^1S_0^{[8]}}
\newcommand{\COt}{\,{}^3S_1^{[8]}}
\newcommand{\COp}{\,{}^3P_J^{[8]}}

\hypersetup{pdftitle={Sensitivity of Jet Multiplicity to J/psi Production Channels at RHIC},pdfauthor={Yu-Tang Wang, Hao Huang, and Yi Yang}}
\title{\bfseries Sensitivity of Jet Multiplicity to $J/\psi$ Production Channels at RHIC}
\author{Yu-Tang Wang$^{1,2}$, \quad Hao Huang$^{1,3}$, \quad and \quad Yi Yang$^{1,2}$\\[0.6em]
\small $^1$Institute of Physics, Academia Sinica, Taipei 11529, Taiwan\\
\small $^2$Department of Physics, National Cheng Kung University, Tainan 70101, Taiwan\\
\small $^3$Université Paris-Saclay, CNRS/IN2P3, IJCLab, 91405 Orsay, France}
\date{26 November 2026}
\begin{document}
\twocolumn[{%
\begin{@twocolumnfalse}
\vspace*{-0.9\baselineskip}
\maketitle
\begin{abstract}
The transverse-momentum spectrum and polarization of the $J/\psi$ constrain its production channels, but different channel mixtures can describe both observables. We examine how much the number of accompanying jets adds to these constraints in proton-proton collisions at RHIC. At fixed $J/\psi$ momentum, mixtures with the same polarization coefficients can give different probabilities for an accompanying jet. At 200~GeV we use the STAR spectrum and polarization to predict the charged-jet counts, then fit all three observables to determine how the allowed fractions change. Adding the preliminary momentum-integrated counts shifts the upper endpoint of the spin-triplet S-wave octet fraction by at most 0.2 percentage points in the two fit prescriptions studied. Interchanging the fitted and predicted observables also exposes a limitation of the production templates. When the spectrum is withheld, even the largest central template ratio of yields in the first two momentum bins falls below the measured central ratio. We give momentum-resolved jet predictions and a spectrum-weighted forecast at 500~GeV. Their sensitivity to direct production depends on how well the contribution from beauty decays is known.
\end{abstract}
\vspace{0.7\baselineskip}
\end{@twocolumnfalse}
}]
\thispagestyle{plain}

\section{Introduction}
\label{sec:intro}

The transverse-momentum spectrum and polarization probe complementary aspects of $J/\psi$ production. At RHIC, the color-evaporation model and a next-to-leading-order calculation in nonrelativistic quantum chromodynamics (NRQCD) both describe the STAR inclusive spectrum at 200~GeV within uncertainties over $4<p_T^{J/\psi}<14\GeV$~\cite{STAR:2018spectrum}. In the STAR dimuon sample at $4<p_T^{J/\psi}<10\GeV$, the helicity-frame coefficients $\lambda_\theta$ and $\lambda_\phi$ are both consistent with zero~\cite{STAR:2020polarization}. At much higher momenta in 13~TeV $pp$ collisions, CMS finds a modest transverse alignment for prompt $\jpsi$, with $\lambda_\theta$ approaching $+0.3$~\cite{CMS:2024polarization}. In NRQCD, several short-distance spin and color states contribute to the yield, and their contributions to the net polarization can compensate~\cite{Bodwin:1994jh,Chao:2012}. Even at fixed $\jpsi$ momentum, different mixtures can give similar spin alignment. Their accompanying jets can still differ if the production channels have different recoil distributions.

Measurements of the surrounding hadrons already show production-dependent differences. At 200~GeV, STAR found that the relative $\jpsi$ yield rises faster than linearly with charged-particle multiplicity at midrapidity~\cite{STAR:2018spectrum}. At 13~TeV, LHCb measured the fraction of a forward jet's transverse momentum carried by the $\jpsi$. In the data, the other jet constituents carried a larger share of the momentum in prompt $\jpsi$ jets than predicted by the PYTHIA calculation used there. The same calculation described the distribution for beauty decays~\cite{LHCb:2017injet}. At 5.02~TeV, ATLAS reconstructed jets without the $\jpsi$ decay signal and measured whether a jet lay near the $\jpsi$. In $pp$ collisions a larger fraction of prompt than nonprompt $\jpsi$ mesons were isolated from these jets~\cite{ATLAS:2026jet}. Theoretical studies have examined both the activity around quarkonium~\cite{Lansberg:2008zm} and the momentum fraction or polarization of a $\jpsi$ inside a jet~\cite{Baumgart:2014,Bain:2017,Kang:2017}.

Event-wide jet counts also probe the recoil opposite the $\jpsi$. At 8~TeV, CMS counted jets above $25\GeV$ within $|\eta_{\rm jet}|<1$. In about 94\% of its two-jet events, one of the two counted jets contained both $\jpsi$ decay muons~\cite{CMS:2020injet}. Here we cluster the $\jpsi$ as a tagged constituent, exclude the entire jet containing it, and count every other accepted jet at any azimuth. A $\jpsi$ with only its own jet therefore enters the zero-jet bin. With this definition we ask how much $N_{\rm jet}$ further constrains the channel fractions allowed by the spectrum and polarization.

We model direct production as a positive mixture of four channels. The color-singlet state is $\CS$ (CS), and the three octet states are $\COs$ (S0), $\COt$ (S1), and $\COp$ (PJ). The P-wave multiplet is grouped. These labels specify the short-distance $c\bar c$ state, which can be populated by different incoming partons. ATLAS found that the beauty fraction rises with $p_T^{J/\psi}$ in its 7 and 8~TeV spectra~\cite{ATLAS:2016spectra}. We include separate templates for beauty and excited-charmonium decays in the inclusive RHIC sample.

At 200~GeV, STAR's preliminary charged-jet counts~\cite{STAR:QM2022jet,STAR:QWG2024jet,STAR:DIS2024} allow us to fit each pair of observables and predict the third. We compare the channel fractions allowed by the spectrum and polarization before and after adding the jet counts. Repeating the jet prediction with measured momentum-bin yields tests its dependence on the modeled spectrum. At 500~GeV we use the measured spectrum to forecast the jet counts for a specified polarization in each bin.

\section{Production fractions and observables}
\label{sec:framework}

\subsection{Channel fractions within each momentum bin}

In bin $b$ of $p_T^{J/\psi}$, let $f_{nb}$ be the share of the direct $\jpsi$ yield from channel $n$, with $f_{nb}\geq0$ and $\sum_n f_{nb}=1$ in each bin. The fractions can differ from bin to bin. For either choice of jet constituents, define the channel response and the mixture probabilities by
\begin{equation}
 \begin{aligned}
 F_{nb}(j)&=P(N_{\rm jet}=j\mid n,\jpsi,b),\\
 a_n^{(b)}&=1-F_{nb}(0),\\
 p_b(j)&=\sum_n f_{nb} F_{nb}(j),\\
 A_{k,b}&=\sum_{j\geq k}p_b(j).
 \end{aligned}
 \label{eq:conditional}
\end{equation}
The normalization of $F_{nb}$ includes zero-jet events. We use $a_n^{(b)}$ as shorthand for the probability that channel $n$ gives at least one accepted jet, so $A_{1,b}=\sum_n f_{nb}a_n^{(b)}$. The sample with at least two jets is a subset of the sample with at least one. Subtracting $A_{2,b}$ from $A_{1,b}$ leaves the probability $p_b(1)$ for exactly one jet. In general, $p_b(j)=A_{j,b}-A_{j+1,b}$ for $j\geq1$ and $p_b(0)=1-A_{1,b}$.

The generated rate of channel $n$ in bin $b$ cancels when constructing $F_{nb}$. For a measured direct yield $Y_b$, the predicted yield in jet category $j$ is $Y_b p_b(j)$. For an inclusive sample, let $r_{H,b}$ and $r_{B,b}$ be the fractions from excited charmonia and beauty hadrons in bin $b$, with responses $F_{Hb}(j)$ and $F_{Bb}(j)$. Then
\begin{equation}
\begin{aligned}
 p(j\mid b)={}&(1-r_{H,b}-r_{B,b})\,p_b(j)\\
 &+r_{H,b}F_{Hb}(j)+r_{B,b}F_{Bb}(j),\\
 p(j)={}&\sum_b w_b\,p(j\mid b),\qquad
 w_b=\frac{Y_b}{\sum_{b'}Y_{b'}} ,
\end{aligned}
 \label{eq:inclusive_conditional}
\end{equation}
where $Y_b$ now denotes the measured inclusive yield. This construction allows independent direct mixtures in each bin. The measured spectrum supplies the bin weights and the spin and jet templates determine the response to those mixtures.

\subsection{Joint templates for momentum and jet multiplicity}

A fit to the spectrum also uses the predicted variation in yield across momentum bins. For that test let $T_n(b,j)$ be the simulated distribution normalized to unity over all $b,j$ in the chosen momentum range. The direct distribution $p_D(b,j)$ and the local fraction $f_{nb}$ follow from the integrated channel fractions $g_n$,
\begin{equation}
\begin{aligned}
 p_D(b,j)&=\sum_n g_n T_n(b,j),\\
 f_{nb}&=\frac{g_n S_{nb}}{\sum_m g_m S_{mb}},\\
 S_{nb}&=\sum_j T_n(b,j).
\end{aligned}
\label{eq:mixture}
\end{equation}
Here $g_n\geq0$ and $\sum_n g_n=1$. A single set of $g_n$ can give different local compositions because $S_{nb}$ has a different momentum dependence for each channel.

For inclusive $\jpsi$, the integrated direct, charmonium feed-down, and beauty fractions are $r_D,r_H,r_B$, with sum one. For unit-normalized daughter templates $T_H$ and $T_B$,
\begin{equation}
 p(b,j)=r_D p_D(b,j)+r_H T_H(b,j)+r_B T_B(b,j).
 \label{eq:inclusive}
\end{equation}
Summing over $j$ gives the normalized momentum shape and summing over $b$ gives the multiplicity shape. The two projections retain different information about the same channel fractions.

\subsection{Helicity-frame polarization}

We work in the helicity frame, with spin projections $m=+1,0,-1$. Let $R_{nb}$ be the unit-trace Born spin-density matrix of direct channel $n$ in momentum bin $b$. The spin-density matrix of the mixture is $\rho_b=\sum_n f_{nb}R_{nb}$. In terms of its two moments~\cite{Faccioli:2010polarization},
\begin{equation}
 \begin{gathered}
 u=\rho_{00},\qquad v=2\operatorname{Re}\rho_{1,-1},\\
 \lambda_\theta=\frac{1-3u}{1+u},\qquad
 \lambda_\phi=\frac{v}{1+u}.
 \end{gathered}
 \label{eq:spin}
\end{equation}
The polar coefficient depends on the longitudinal probability $u$, and the azimuthal coefficient probes interference between the two transverse helicities. We leave the mixed polar--azimuthal coefficient $\lambda_{\theta\phi}$ unconstrained. For a wider momentum range or an inclusive sample, the matrices are weighted by their respective yields before Eq.~\eqref{eq:spin} is evaluated. In the inclusive calculations we profile the combined spin of the two secondary sources over $0\le u\le1$ and $|v|\le1-u$, the physical region for these moments.

\section{Simulation and data}
\label{sec:inputs}

\subsection{Event generation and spin templates}

We use PYTHIA~8.318 with the Monash tune, initial- and final-state radiation (ISR and FSR), and multiparton interactions (MPI)~\cite{Bierlich:2022,Skands:2014}. Direct production uses $2\to2$ hard processes. The baseline combines disjoint intervals of hard-scattering momentum $\hat p_T\geq1\GeV$ using their generated cross sections. The production of new quarkonia in showers is disabled. Radiation from an already produced octet is enabled with \texttt{OniaShower:octetSplit=1} and \texttt{Onia:massSplit=0.2}, and the decay-length setting is \texttt{ParticleDecays:tau0Max=10.0}~mm~\cite{Pythia:manualRadiation,Cooke:2023}.

For the momentum-resolved direct sample we compare the baseline with three variants. One replaces the massive-gluon octet kernel by the heavy-quark radiation kernel rescaled by $C_A/C_F=9/4$. Here $C_A=3$ and $C_F=4/3$ are the adjoint and fundamental quadratic Casimirs of QCD. The other two multiply the shower renormalization scale by $1/2$ or $2$ for ISR and FSR together. The four settings use a total of 839.6 million independent direct-production events across the two energies. The octet-kernel variation shares the baseline CS sample. The spin-density matrices come from HELAC-Onia at Born level with CTEQ6L1 parton distributions, $m_c=1.5\GeV$, and factorization and renormalization scales $\sqrt{4m_c^2+(p_T^{J/\psi})^2}$~\cite{Shao:2015helac}. We integrate the Born spin matrices in each momentum bin and use the same matrices for all four shower settings. In particular $R_{\rm S0}=I/3$.

The simulated inclusive sample contains $\jpsi$ from unforced decays of excited charmonia and beauty hadrons. The excited-charmonium sample combines $\chi_{cJ}$ and $\psi(2S)$ production at a fixed positive matrix-element point. The beauty sample uses $gg,q\bar q\to b\bar b$. At each energy we generate 550,000 excited-charmonium and 500,000 beauty hard events in five $\hat p_T$ intervals 1--2, 2--4, 4--8, 8--16, and at least 16~GeV. We weight each interval by its generated cross section divided by the sum of its event weights. The MC covariance retains correlations between candidates from the same event.

At 200~GeV the daughter templates below $p_T^{J/\psi}=4\GeV$ use one million additional hard events, 500,000 per source across the same five $\hat p_T$ intervals.

We used OpenAI ChatGPT and Codex to assist with implementing and checking the analysis code.

\subsection{Jet definitions and fiducial selection}

We use $|y^{J/\psi}|<0.4$ to match the STAR charged-jet selection~\cite{STAR:QWG2024jet}. The integrated jet comparison uses $p_T^{J/\psi}<10\GeV$. For the momentum-resolved comparison we choose $4\leq p_T^{J/\psi}<10\GeV$, the upper bin of the STAR dimuon polarization measurement at $|y^{J/\psi}|<0.5$~\cite{STAR:2020polarization}. In our 4--10~GeV fiducial window, the 200~GeV PYTHIA samples yield 13,509 daughter $\jpsi$ candidates from 550,000 generated excited-charmonium events and 1,289 from 500,000 generated beauty events. For the same sample sizes at 500~GeV, the corresponding candidate counts are 9,929 and 892.

We cluster the $\jpsi$ as a single tagged constituent with the other particles using anti-$k_t$ ($R=0.4$) in FastJet~\cite{Cacciari:2008,Cacciari:2012}. For dimuons, the pair replaces both muon tracks with their summed four-momentum, entered once. In the simulation the stable $\jpsi$ supplies this four-momentum. We use four-vector recombination, identify the tagged jet by constituent membership, exclude that entire jet, and count all other jets with $p_T^{\rm jet}\geq3\GeV$ and $|y_{\rm jet}|<0.6$. The count includes all azimuths relative to the $\jpsi$. Other constituents must have $p_T>0.2\GeV$ and $|\eta|<1$. Charged jets use charged particles in addition to the tag. Full jets also include visible neutral particles.

Each selected $\jpsi$ contributes one multiplicity entry. If no accompanying jet passes the cuts, the candidate enters the zero-jet bin and remains in the normalization. Thus $A_1$ is the fraction of all selected $\jpsi$ in the specified momentum bin with at least one accompanying jet.

We use this prescription for all direct and secondary templates and for the independent 2.4-million-event recoil and MPI control sample.

For the 200~GeV inclusive comparison, this selection gives reference generator charmonium and beauty fractions of 23.62\% and 1.25\% for 0--10~GeV, and 25.30\% and 3.64\% for 4--10~GeV. Within a given fit we hold the chosen source fractions fixed. The four direct-channel weights divide the remaining direct yield.

\subsection{Experimental inputs at 200 GeV}

The STAR inclusive spectrum is measured in $|y^{J/\psi}|<1$~\cite{STAR:2018spectrum}. We approximate each bin integral using its tabulated invariant cross section times $2\pi p_{T,b}\Delta p_{T,b}$. We combine bins up to $8\GeV$ into one-GeV bins and keep 8--10~GeV together. The fit assumes the normalized momentum shape transfers to $|y^{J/\psi}|<0.4$. The PHENIX spectrum provides an independent shape check within its own $|y^{J/\psi}|<0.35$ window~\cite{PHENIX:2012psi}.

For polarization we use the STAR dimuon results in Table~VI of Ref.~\cite{STAR:2020polarization}. They give $\lambda_\theta=0.12\pm0.42\pm0.17$ and $\lambda_\phi=-0.07\pm0.19\pm0.07$ for $4<p_T^{J/\psi}<10\GeV$ and $|y^{J/\psi}|<0.5$ in the helicity frame. We use these results in the central $|y^{J/\psi}|<0.4$ benchmark and add the statistical and systematic errors in quadrature. The constraint from these data applies to that momentum range.

The STAR preliminary charged-jet result integrates over $p_T^{J/\psi}<10\GeV$ in $|y^{J/\psi}|<0.4$. The public points predate the tagged-jet prescription used in our simulation. We digitize the cross-section points for the bins labeled 0, 1, and 2 in the $R=0.4$ panel on slide~14 of Ref.~\cite{STAR:QWG2024jet}. We propagate the statistical and systematic uncertainties read from the plot, using the marker size as an upper estimate of the unresolved statistical error in the zero-jet bin. A common luminosity scale cancels upon normalization. The primary calculation uses exactly two jets for the last category and checks the alternative inclusive bin with at least two jets. A joint distribution in momentum and jet counts would retain the event-level correlations.

\section{Sensitivity at fixed \texorpdfstring{$J/\psi$}{J/psi} momentum}
\label{sec:local}

\subsection{Spin and jet activity of direct-channel mixtures}

\begin{figure*}[t]\centering
\includegraphics[width=.935\linewidth]{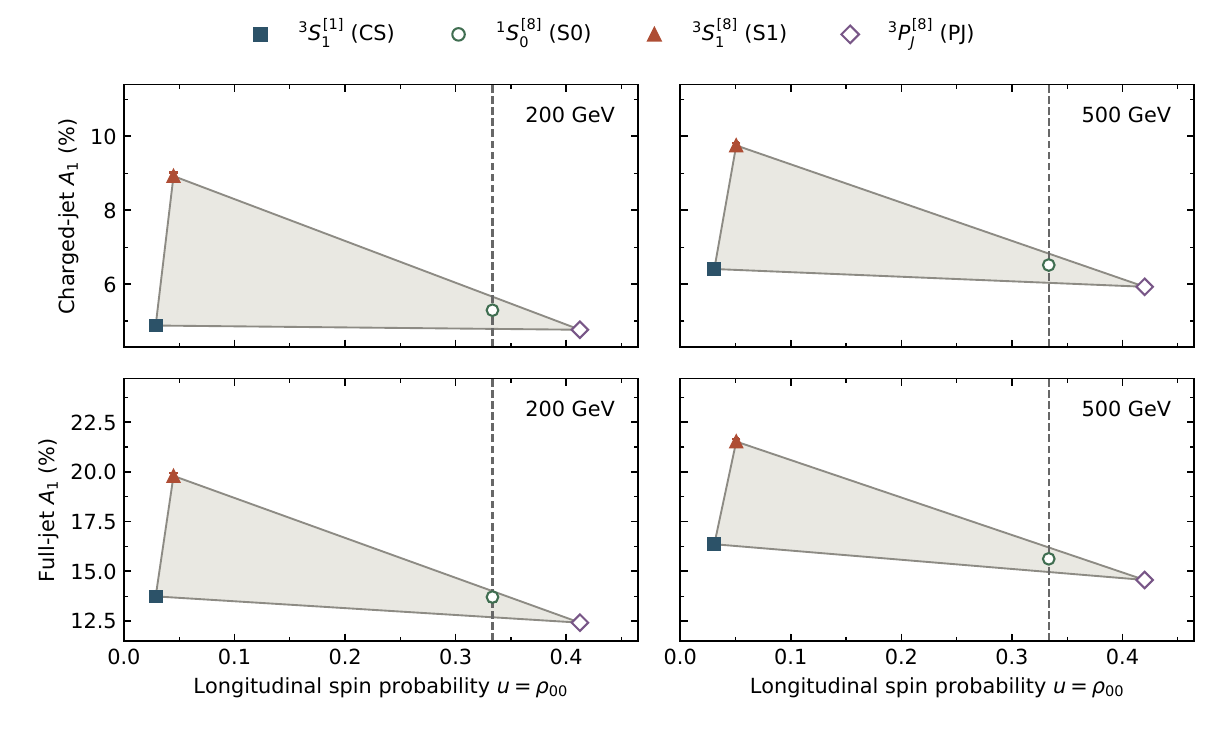}
\caption{Baseline response of each direct channel for $6<p_T^{J/\psi}<7\GeV$ at 200 and 500~GeV. A point gives the longitudinal spin probability and the percentage of all signals in that channel with at least one jet. The upper panels use charged jets and the lower panels use full jets. The shading shows the combinations attainable by positive mixtures. The vertical dashed line selects $\lambda_\theta=0$. The vertical bars give jet-template MC standard errors, and the horizontal bars give spin integration errors.}
\label{fig:response}
\end{figure*}

Figure~\ref{fig:response} compares the direct channels in the representative bin $6<p_T^{J/\psi}<7\GeV$. Each marker represents one pure channel. The shaded region contains all pairs of longitudinal spin probability and jet probability obtained by mixing the four channels with nonnegative fractions that sum to one. CS and S1 both have small longitudinal probabilities, yet their jet probabilities differ. S0 and PJ have similar jet probabilities, while their longitudinal probabilities differ.

With either jet definition, S1 has the highest activity in this bin for each of the four common shower settings. The full-jet probabilities also vary more across shower settings. The following benchmarks use charged jets for comparison with the STAR observable.

The independent recoil control helps explain which charged jets carry this contrast. For S1 the probability of a jet in the hemisphere opposite the $\jpsi$ is $(7.6\pm0.7)\%$ at 200~GeV and $(7.9\pm0.8)\%$ at 500~GeV. After restricting S0 and S1 to $gg$ and $qg$ and reweighting both to the S0 initial-state fractions, we still find an S1 excess of $(3.4\pm1.0)$ and $(4.2\pm1.5)$ percentage points. This control locates most S1 tags in the opposite hemisphere and establishes that the excess survives this initial-state reweighting. The accepted recoil can depend on both hard kinematics and showering.

To test whether the jet probabilities can still differ when both spin coefficients are fixed, we select two mixtures in each momentum bin with $\lambda_\theta=0$ and a common $\lambda_\phi$. We choose the common coefficient and the two mixtures that maximize the baseline charged-jet difference. We subsequently compare those same mixtures across shower settings. These are direct-signal benchmarks at fixed momentum.

\begin{table}[htb]\centering\small
\begin{tabular}{@{}llrrrrr@{}}
\toprule
$\sqrt{s}$ [GeV] & Mixture & CS & S0 & S1 & PJ & $A_1$ [\%]\\
\midrule
200 & I  & 16.8 & 18.7 & 0 & 64.5 & 4.88\\
    & II & 0 & 0 & 21.5 & 78.5 & 5.66\\
\midrule
500 & I  & 15.2 & 32.0 & 0 & 52.8 & 6.19\\
    & II & 0 & 0 & 23.5 & 76.5 & 6.83\\
\bottomrule
\end{tabular}
\caption{Worked example in the single bin $6<p_T^{J/\psi}<7\GeV$. The four channel columns give percentages of the direct $\jpsi$ yield. The last column gives the percentage of those signals with at least one accepted charged jet. At $\lambda_\theta=0$ the two mixtures minimize and maximize baseline $A_1$ at the common $\lambda_\phi$ chosen to maximize their separation ($-0.0132$ at 200~GeV and $-0.0086$ at 500~GeV).}
\label{tab:examples}
\end{table}

Interpolating between I and II in Table~\ref{tab:examples} keeps both decay coefficients fixed and varies baseline $A_1$ linearly. For both energies, mixture II gives a larger central activity than mixture I at each common shower setting. A measured activity must be compared with the full response range of each mixture.

\begin{figure*}[t]\centering
\includegraphics[width=.935\linewidth]{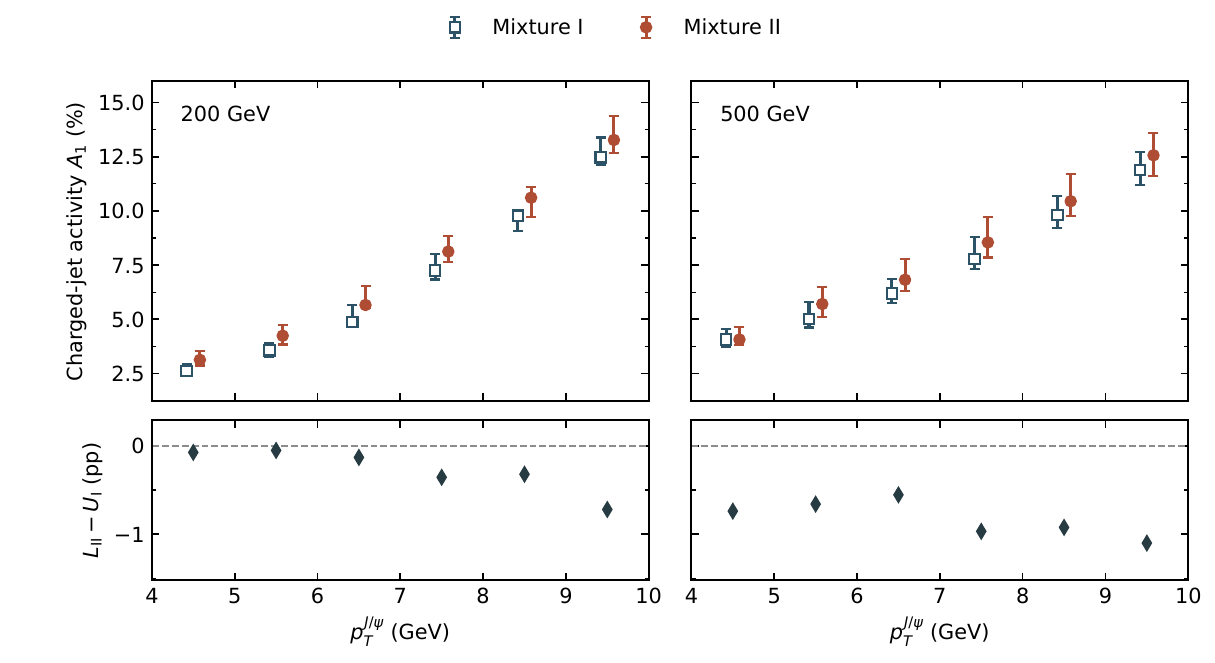}
\caption{Direct mixtures with equal $\lambda_\theta=0$ and equal $\lambda_\phi$, selected independently in each momentum bin. The top panels show their baseline activities and the ranges from the four shower settings with the MC sampling margin. The bottom panels give the lower edge for II minus the upper edge for I, in percentage points. Positive values indicate disjoint prediction ranges for these two mixtures.}
\label{fig:pairs}
\end{figure*}

Write $A_1^{(r)}(f)=\sum_n f_n a_n^{(r)}$ for an activity at fixed momentum and shower setting $r$. For each setting we take a sampling allowance $\xi_r$ equal to the largest conditional MC standard error over the fraction set being scanned. The activity range $[L,U]$ for mixture $f$ is
\begin{equation}
 \begin{aligned}
 L(f)&=\min_r\big[A_1^{(r)}(f)-\xi_r\big],\\
 U(f)&=\max_r\big[A_1^{(r)}(f)+\xi_r\big].
 \end{aligned}
\end{equation}
The sampling allowances and the shower variations define a model sensitivity range.

Figure~\ref{fig:pairs} extends this comparison to six momentum bins at each energy. The separation of the baseline activities varies with momentum. Once the shower and MC sampling ranges are included, the two equal-polarization benchmarks overlap in every bin at both energies.

\subsection{Channel-fraction bounds from jet activity}

\begin{figure*}[t]\centering
\includegraphics[width=.935\linewidth]{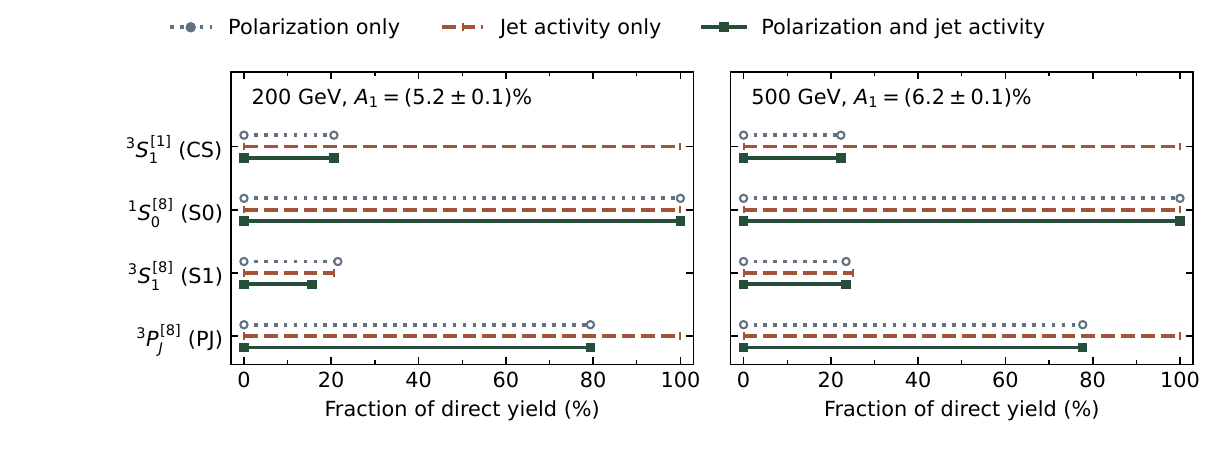}
\caption{Direct-channel fraction ranges in $6<p_T^{J/\psi}<7\GeV$ for hypothetical measurements of the at-least-one-charged-jet probability $A_1$. Polarization selects the Born slice $\lambda_\theta=0$ and leaves $\lambda_\phi$ free. Each segment shows the possible fractions of one channel after varying the other three. Comparing polarization alone with the joint constraint shows how much the chosen jet input can tighten the range. The activity constraints include shower and sampling variations.}
\label{fig:localbounds}
\end{figure*}

We invert the activity response for example measurements at the two energies in Fig.~\ref{fig:localbounds}. For each channel we find its minimum and maximum share among the mixtures whose activity range intersects the example measurement interval. The calculation is repeated with and without the angular constraint. At 200~GeV, a fraction of $(5.2\pm0.1)\%$ of direct signals with an accepted jet would bound S1 to 20.6\% using jet activity alone. Adding this input to the Born polar constraint reduces the S1 maximum from 21.5\% to 15.6\%. The corresponding three maxima at 500~GeV and $A_1=(6.2\pm0.1)\%$ are 25.1\% from activity, 23.5\% from polarization and 23.5\% jointly. At this energy the jet input leaves the polarization bound unchanged. On the $\lambda_\theta=0$ slice, a nonzero CS or S1 share requires a compensating PJ contribution. S0 already lies on that slice. Removing the angular input tests which limits survive without the Born spin matrices.

In the 200~GeV example, the baseline central activities for pure CS, S0, and PJ are 4.88\%, 5.29\%, and 4.77\%, respectively. Only pure S0 lies within the illustrative interval of 5.1--5.3\%. With the shower and sampling ranges included, pure CS and pure PJ are also compatible. At twice the nominal shower scale every mixture along the CS--S0 segment fits the interval with the sampling allowance, so the total octet share can take any value from zero to one using jet activity alone. The S1 bound follows from its higher jet probability. This is a constraint on a particular octet contribution. It explains why sensitivity to a single channel and sensitivity to the total octet yield can differ substantially.

\section{Predicting each observable from the other two at 200 GeV}
\label{sec:datafit}

\begin{figure*}[t]\centering
\includegraphics[width=.97\linewidth]{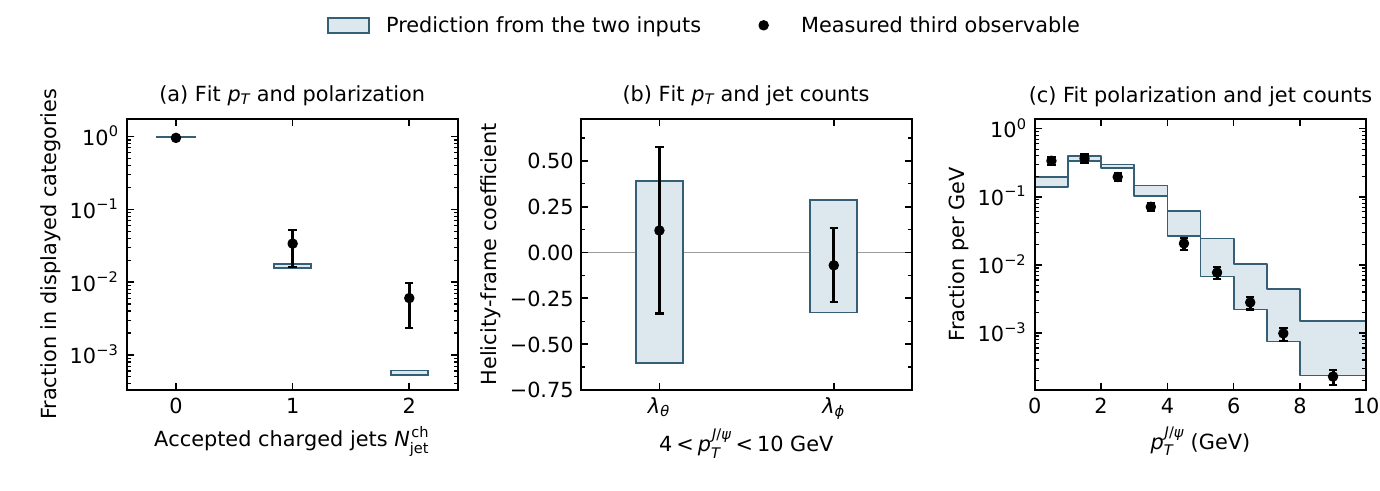}
\caption{Three predictions at 200~GeV. In each panel the title names the observables used in the fit, and the horizontal axis identifies the held-out observable. The shaded regions give marginal ranges at the common fit tolerance, including the physical secondary-spin range in panel (b). The points give the corresponding STAR measurements with marginal errors. The charged-jet distribution is normalized over the three displayed bins from the digitized preliminary plot. The polarization coefficients are integrated over 4--10~GeV, and the spectrum and charged-jet distribution cover 0--10~GeV.}
\label{fig:fits}
\end{figure*}

The fixed-momentum comparison isolates direct production. We now turn to the inclusive STAR measurements, where both feed-down and beauty decays contribute. Figure~\ref{fig:fits} compares each of the three measured observables with the prediction from fitting the other two. For each pair, one set of integrated direct fractions fixes the mixture in every momentum bin through Eq.~\eqref{eq:mixture}. The spectrum and jet counts span 0--10~GeV, while the angular coefficients constrain the 4--10~GeV part of the same mixture. Our primary fit uses normalized shapes for the spectrum and jet counts. We propagate the yield covariance to the normalized fractions by linearizing about the measured yields and omit one redundant entry per distribution. Each fit uses the relevant submatrix. We include the Monte Carlo covariance between the two projections at the reference composition. In the primary calculation, point-to-point data errors are independent before normalization. For the spectrum, we approximate the normalization covariance by two independent scale shifts, 10\% below $4\GeV$ and 8.5\% at and above $4\GeV$. The comparison assumes zero covariance between distinct experimental inputs and between angular coefficients, since these covariances are unavailable. We use the quadratic discrepancy $Q=\mathbf{d}^{T}C^{-1}\mathbf{d}$ for the measured-minus-predicted vector $\mathbf{d}$ and its covariance $C$. When polarization enters the fit, we profile the combined spin of the secondary sources over its physical range.

For each pair we retain the mixtures within $Q-Q_{\min}\leq4$ and compute their predictions for the omitted observable. The value 4 sets the scale of a two-standard-deviation displacement for one ideal Gaussian fit parameter. Here the mixture weights have positivity boundaries and the spectrum can be poorly described. We use the cut as a common sensitivity tolerance without a calibrated confidence level.

\subsection{Pairwise fits to the STAR data}

The mixtures fitted to the spectrum and polarization predict fewer two-jet events relative to one-jet events than the preliminary central values in Fig.~\ref{fig:fits}(a). The predicted ratio is 0.032--0.036, compared with approximately $0.18\pm0.12$. The difference is about 1.2 experimental standard errors under our diagonal error prescription. Counting at least two jets in the last category gives a similar comparison.

In Fig.~\ref{fig:fits}(b), mixtures fitted to the spectrum and jet counts can reproduce both measured angular coefficients simultaneously when the secondary spin varies over its physical domain. The allowed angular ranges are broad. These two shapes therefore provide only a weak prediction of the polarization in 4--10~GeV.

When the spectrum is withheld, the mixtures fitted to polarization and jet counts underpredict the yield in the lowest momentum bin, as shown in Fig.~\ref{fig:fits}(c). The predicted ratio $R_{01}$ of the yields in 0--1 and 1--2~GeV is 0.397--0.492, compared with the transferred STAR value of $0.93\pm0.25$ under our error prescription. For any positive mixture, $R_{01}$ is a weighted average of the individual source ratios with weights proportional to each source's yield in the denominator bin. The largest central ratio among the six source templates is 0.51. Adjusting the channel fractions alone therefore cannot reproduce the STAR central ratio with these regulated PYTHIA templates.

We also calculate the ratio from PHENIX Table~VII in $|y^{J/\psi}|<0.35$ and apply the same cut to simulated daughters~\cite{PHENIX:2012psi}. The data give $0.629\pm0.029$ using midpoint integration, point-to-point errors, and a common shift from the quoted correlated errors. The corresponding central template ratios span 0.391--0.504. This comparison with matched rapidity gives the same direction of the discrepancy in the lowest momentum bin.

\subsection{Channel constraints with and without jet counts}

\begin{figure*}[t]\centering
\includegraphics[width=.935\linewidth]{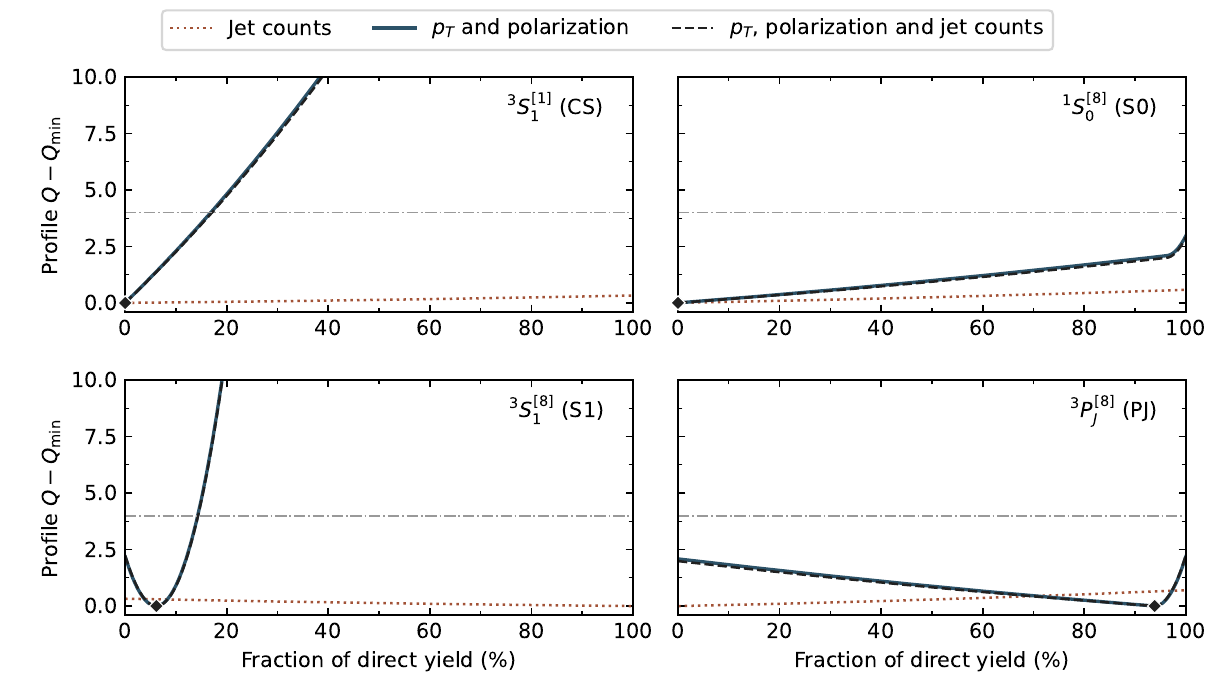}
\caption{Profiles of the integrated direct fractions $g_n$ from fits to normalized distributions at 200~GeV and $p_T^{J/\psi}<10\GeV$. At each horizontal coordinate, the other fractions are varied to minimize $Q$, then the minimum over all fractions is subtracted. The horizontal line at $\Delta Q=4$ marks the common tolerance. The dotted curve uses jets alone. The solid curve uses the spectrum and polarization. The dashed curve adds jet counts. The four diamonds mark one combined minimum. In the CS panel the minimum is at zero, and the dashed curve remains within the tolerance up to a 17.1\% share.}
\label{fig:profiles}
\end{figure*}

The 0--10~GeV fit to the spectrum and polarization gives $Q_{\min}=27.99$, entirely from the spectrum at the minimum. For a diagnostic refit we normalize the spectrum and templates and compute the reference source fractions over 1--10~GeV, keeping the same polarization input. The minimum drops to 8.02, consistent with the lowest-bin deficit discussed above. The public STAR jet projection integrates over the full 0--10~GeV interval, so we use that interval to assess the incremental effect of the jet data in the presence of the spectrum discrepancy.

At the reference secondary rates the combined fit to normalized shapes and polarization has its minimum at $g_{\rm CS}=g_{\rm S0}=0$, with S1 and PJ supplying the direct yield. Pure S0 and pure PJ each lie within the adopted tolerance. Thus the location of the minimum gives little guidance on their relative contributions.

Figure~\ref{fig:profiles} shows the incremental effect of jet counts on the integrated fractions. At $\Delta Q=4$ the S1 upper endpoint moves from 14.30\% to 14.37\% when jets are added. Small increases are possible because each curve uses its own minimum. The CS and S1 upper endpoints move by less than 0.4 percentage points at each of the three tolerances $\Delta Q=2$, 4, and 6 in the normalized-shape fit.

Fitting the original yields with separate free normalizations and fixed MC templates instead gives a combined minimum dominated by S0. The lower bound on $g_{\rm S0}+g_{\rm PJ}$ in the combined fit falls from 76.0\% for normalized shapes to 9.0\% for the original yields. Within the yield fit, adding jets moves the S1 upper endpoint only from 27.6\% to 27.8\%. The two treatments thus agree on how little the integrated jet counts add to this bound, even though the inferred composition depends strongly on the spectrum fit.

The shape fit linearizes the normalization at the measured yields. A free normalization rescales the bins in the proportions predicted by each mixture. For fixed templates and the same yield covariance, expanding at the predicted shape and measured total reproduces the profiled yield term in $Q$. Switching the spectrum and the jet counts to original yields one at a time, with fixed templates throughout, shows that the spectrum treatment accounts for most of the change.

\subsection{Conditioning on the measured spectrum}

To test whether the modeled spectrum accounts for the low predicted two-jet to one-jet ratio, we fix each momentum-bin yield to the measured spectrum and allow the four direct fractions to vary independently between bins. The inclusive jet distribution follows by multiplying the per-bin jet probabilities by their bin yields and summing, with the reference feed-down and beauty fractions in each bin. This relaxes the generated spectrum shape even when it differs from data.

The full range of the two-jet to one-jet ratio across these mixtures is 0.022--0.054 for exactly two jets, or 0.023--0.055 if the last category includes all higher counts. The public plot gives approximately $0.18\pm0.12$ with diagonal experimental errors. Its central value favors more multijet activity than the templates supply, while the uncertainty limits the sensitivity to direct composition. The ratio avoids the large uncertainty from the zero-jet normalization. Even when each momentum bin can choose its own fractions, the largest predicted ratio lies about one experimental standard error below the preliminary central value.

\section{Jet sensitivity to the production mixture in \texorpdfstring{$4<p_T^{J/\psi}<10$}{4 < pT < 10} GeV}
\label{sec:prediction}

About 3.25\% of the 0--10~GeV yield in the measured STAR spectrum at $|y^{J/\psi}|<1$ lies in 4--10~GeV. The present jet distribution is therefore dominated by lower momenta. We fit the spectrum and polarization within 4--10~GeV, where the minimum is $Q=3.80$ for the four independent spectrum entries and two angular coefficients, and predict the jet multiplicity in exactly that window. In this fit to the normalized spectrum and polarization, the full range $0\leq g_{\rm CS}\leq1$ lies within $\Delta Q\leq4$.

\begin{table*}[tb]\centering\small
\begin{tabular}{lrrrr}
\toprule
$p_T^{J/\psi}$ [GeV] & Baseline $A_1$ [\%] & MC error [pp] & $A_1$ across shower settings [\%] & Baseline $A_2$ [\%]\\
\midrule
4--5 & 2.98--3.72 & 0.15 & 2.90--3.99 & 0.16--0.21\\
5--6 & 5.35--6.53 & 0.37 & 5.30--6.99 & 0.22--0.29\\
6--7 & 8.05--9.38 & 0.67 & 8.00--9.71 & 0.25--0.38\\
7--8 & 10.22--12.06 & 0.75 & 10.05--12.55 & 1.08--1.25\\
8--9 & 16.06--17.02 & 1.40 & 15.79--17.61 & 1.75--2.18\\
9--10 & 21.61--22.41 & 3.11 & 20.97--22.76 & 1.99--2.29\\
\midrule
4--10 & 4.05--5.36 & 0.13 & 4.00--5.69 & 0.22--0.30\\
\bottomrule
\end{tabular}\space\space
\caption{Inclusive charged-jet predictions at 200~GeV from the spectrum and polarization in 4--10~GeV. $A_k$ is the probability for at least $k$ accepted charged jets. Each bin uses the same fit family. Probabilities are expressed in percent. The MC column gives the larger baseline standard error at the two $A_1$ endpoints in percentage points (pp), conditional on the fitted weights. The direct-response column changes the direct conditional activity over the four shower settings, at fixed momentum and spin templates and secondary sources.}
\label{tab:heldout}
\end{table*}

Table~\ref{tab:heldout} gives the predictions in six momentum bins and for the full 4--10~GeV window. The larger sampling errors in the last bins limit the precision there. These inclusive rates contain the jet activity of feed-down and beauty events in addition to the direct-channel responses in Fig.~\ref{fig:response}.

To isolate the sensitivity to the direct mixture, we first fix the secondary fractions at their generator reference values. Compositions with S1 supplying 10\% or 50\% of the direct 4--10~GeV yield both fit the spectrum and polarization within $\Delta Q\leq4$ (Table~\ref{tab:sources}). Their nominal $A_1$ ranges across shower settings are separated by 0.35 percentage points, calculated from the unrounded endpoints. The baseline MC standard error is about 0.13 percentage points at each endpoint. At a common shower setting the predictions share MC templates, so the sampling uncertainty of their difference depends on their covariance.

\begin{table*}[tb]\centering\small
\begin{tabular}{lrrrrrr}
\toprule
Scenario & $r_H$ [\%] & $r_B$ [\%] & $g_{\rm S1}$ [\%] & $Q_{\min}$ & $A_1$ [\%] & $A_2$ [\%]\\
\midrule
Reference & 25.30 & 3.64 & free & 3.80 & 4.00--5.69 & 0.22--0.34\\
Charmonium varied & 42.00 & 3.64 & free & 4.05 & 4.33--5.71 & 0.26--0.35\\
Beauty varied & 25.30 & 10.00 & free & 4.04 & 4.71--6.19 & 0.31--0.41\\
 & 25.30 & 15.00 & free & 4.22 & 5.39--6.60 & 0.39--0.47\\
 & 25.30 & 25.00 & free & 5.36 & 6.74--7.54 & 0.55--0.60\\
Both varied & 42.00 & 15.00 & free & 4.98 & 5.86--6.67 & 0.44--0.49\\
\midrule
Composition tests & 25.30 & 3.64 & 10 & 4.91 & 4.08--4.59 & 0.23--0.26\\
 & 25.30 & 3.64 & 50 & 5.76 & 4.95--5.51 & 0.27--0.33\\
with $r_B=10\%$ & 25.30 & 10.00 & 10 & 4.15 & 4.91--5.39 & 0.32--0.35\\
\bottomrule
\end{tabular}\space\space
\caption{Source-rate tests of the inclusive 4--10~GeV prediction at 200~GeV. The source fractions $r_H,r_B$ refer to the inclusive yield and $g_{\rm S1}$ to the direct yield. The fractions and the jet probabilities $A_1,A_2$ are expressed in percent. The reference row uses generator rates. The upper rows fit $g_{\rm S1}$. Each source choice is refitted to the spectrum and polarization. In the last three rows the shown S1 fraction is held fixed, and the jet ranges use the tolerance from the corresponding fit with all four direct weights free to vary. The jet columns give nominal ranges over mixtures and direct shower settings. MC sampling errors on the jet predictions are quoted separately for the benchmarks.}
\label{tab:sources}
\end{table*}

We next vary the secondary yields to see whether the separation persists in an inclusive sample. PHENIX reports a combined charmonium feed-down fraction of $(42\pm9)\%$ at midrapidity~\cite{PHENIX:2012psi}. STAR extracted beauty fractions of 10--25\% across $p_T^{J/\psi}=4$--$12\GeV$ from $\jpsi$--hadron correlations using PYTHIA to model the near-side correlation shapes~\cite{STAR:2013beauty}. These estimates motivate the source-rate variations in Table~\ref{tab:sources}, since their selections differ from our 4--10~GeV benchmark. In each scenario we refit the same spectrum and polarization, retaining the simulated shapes and charged-jet responses of the two secondary sources.

Beauty daughters have a simulated probability of 16.13\% for at least one jet in this window. Raising the beauty fraction to 10\% makes the $A_1$ range for 10\% direct S1 overlap that for 50\% direct S1 at the reference beauty rate (Table~\ref{tab:sources}). Their $A_2$ ranges also overlap slightly. How well the two rates together can distinguish these cases depends on their experimental and template covariances. The simulated secondary jet responses and the transfer in rapidity remain additional sources of uncertainty for an inclusive measurement.

\section{Charged-jet forecast at 500 GeV}
\label{sec:forecast}

At 500~GeV, we take the inclusive $J/\psi\to e^+e^-$ spectrum in Table~II of Ref.~\cite{STAR:2019spectrum} and use its unpolarized central values to set $w_b$ in Eq.~\eqref{eq:inclusive_conditional} for six one-GeV bins between 4 and 10~GeV. We multiply the invariant cross section in each published bin by its $2\pi p_T\Delta p_T$ factor before aggregating to these bins. The measured shape at $|y^{J/\psi}|<1$ is transferred to the jet selection at $|y^{J/\psi}|<0.4$. The first two one-GeV bins carry about 82\% of the yield in this window.

We set the inclusive helicity-frame coefficients $\lambda_\theta=\lambda_\phi=0$ in each bin as a benchmark. STAR has reported a separate 500~GeV polarization measurement over $5<p_T^{J/\psi}<16\GeV$~\cite{STAR:2015spin500}. We allow the direct fractions to vary independently across the six bins, profile the combined secondary spin in its physical region, and use the generator reference fractions for the secondary yields in each bin. Weighted by $w_b$, the charmonium and beauty fractions are 27.8\% and 5.9\%, respectively.

\begin{figure}[tb]\centering
\includegraphics[width=\linewidth]{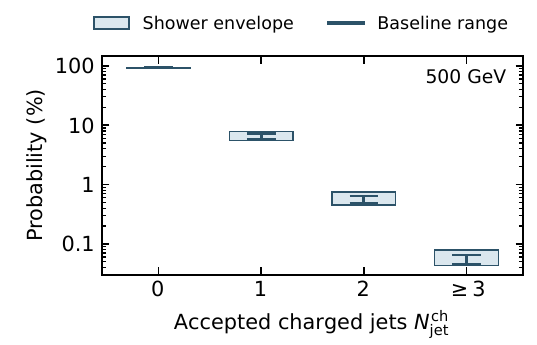}
\caption{Nominal charged-jet multiplicity ranges at 500~GeV for inclusive $\jpsi$ with $4<p_T^{J/\psi}<10\GeV$, $\lambda_\theta=\lambda_\phi=0$ per momentum bin, and fixed reference secondary rates. Bars span the allowed mixtures for the baseline templates. Boxes also include shower variation. Table~\ref{tab:forecast} lists baseline MC errors.}
\label{fig:forecast}
\end{figure}

\begin{table}[tb]\centering\small
\begin{tabular}{lccc}
\toprule
$N_{\rm jet}^{\rm ch}$ & Baseline (\%) & Shower (\%) & MC (pp)\\
\midrule
$0$ & 92.05--93.75 & 91.42--93.93 & 0.19\\
$1$ & 5.72--7.25 & 5.57--7.76 & 0.18\\
$2$ & 0.478--0.636 & 0.455--0.744 & 0.055\\
$\geq3$ & 0.045--0.065 & 0.043--0.078 & 0.032\\
\midrule
$A_1$ & 6.25--7.95 & 6.07--8.58 & 0.19\\
\bottomrule
\end{tabular}\space\space
\caption{Inclusive charged-jet multiplicity at 500~GeV for $4<p_T^{J/\psi}<10\GeV$ with $\lambda_\theta=\lambda_\phi=0$ in each bin. The two ranges scan the compatible compositions for the baseline templates and for all four direct shower settings, at fixed reference secondary rates. The last column gives the larger baseline endpoint MC standard error in percentage points, at fixed fractions and spectrum weights.}
\label{tab:forecast}
\end{table}

Figure~\ref{fig:forecast} shows the predictions for zero, one, two, and at least three jets. Table~\ref{tab:forecast} also lists $A_1$. For each category and common shower setting we minimize and maximize its weighted sum over the allowed mixtures. The extrema for different rows can come from different mixtures. The nominal $A_1$ range is 6.07--8.58\% when varying the direct shower. Propagating the quoted point-to-point spectrum errors as independent gives another 0.09--0.11 percentage points at the baseline $A_1$ endpoints. The rapidity transfer and assumptions on secondary rates and responses also enter the prediction.

Secondary sources make up about 34\% of the spectrum-weighted yield. At the reference rates each pure direct channel can satisfy both zero angular coefficients in every bin above 5~GeV by adjusting the secondary spin. In the 4--5~GeV bin the pure CS and S1 compositions are excluded. For pure S1 the required secondary moments would be $(u,v)\simeq(0.87,-0.22)$, outside $|v|\leq1-u$. Removing both angular constraints leaves the lower $A_1$ endpoint unchanged and raises the upper endpoint by 0.06 percentage points in the baseline and 0.07 percentage points across shower settings. With the secondary spin free within its physical domain, the two inclusive angular constraints thus leave nearly the full range of $A_1$ accessible.

We test a higher beauty yield by multiplying the beauty-to-direct ratio by the same factor in each bin while holding the charmonium fractions, spectrum weights, and component jet responses fixed. A factor of 3.02 gives a spectrum-weighted beauty fraction of 15\%. Keeping $\lambda_\theta=\lambda_\phi=0$, the $A_1$ range is 7.48--8.99\% with the baseline templates and 7.33--9.53\% across the four shower settings. The baseline endpoint MC errors are about 0.32 percentage points.

\section{Conclusions}
\begingroup\linespread{0.94}\selectfont

Starting from the channel mixtures allowed by the $J/\psi$ spectrum and polarization at 200~GeV, we find that adding the preliminary momentum-integrated jet counts changes the upper endpoint of the direct $\COt$ fraction by at most 0.2 percentage points at $\Delta Q=4$ in the two fit prescriptions studied. The preferred composition depends strongly on how the spectrum enters the fit, and the relative contributions of $\COs$ and $\COp$ remain unresolved. At fixed momentum, mixtures with the same $\lambda_\theta$ and $\lambda_\phi$ can nevertheless predict different jet probabilities. The shower and sampling ranges of the equal-polarization benchmarks overlap in all six bins.

Predicting each observable from the other two reveals where the common production description is least successful. Fitting the spectrum and jet counts accommodates the measured polarization, although the angular prediction is broad. Fitting the spectrum and polarization gives less multijet activity than the preliminary central values, with limited experimental precision. The most informative test is the spectrum predicted from polarization and jets. Every central source template gives too little yield in the lowest momentum bin relative to the next bin to reproduce the measured central ratio. This identifies a limitation of the modeled momentum dependence that adjusting the direct fractions cannot remove.

A measurement of accompanying-jet multiplicity in $J/\psi$ momentum bins, including candidates with zero jets, can test the jet response with the bin yields fixed to the measured spectrum. The inclusive interpretation also requires information on beauty decays. In the 4--10~GeV window at 200~GeV, varying the beauty rate can bring the jet probabilities of two very different direct mixtures into agreement. At 500~GeV the spectrum-weighted forecast likewise depends more strongly on the tested change in beauty fraction than on imposing zero inclusive polarization with freely varying physical secondary spin.

At the LHC, fits to the displacement of the dimuon vertex have separated prompt $J/\psi$ production from beauty-hadron decays~\cite{CMS:2011prompt,LHCb:2017injet}. Applying such fits in each $(p_T^{J/\psi},N_{\rm jet})$ bin, including zero jets, would give separate prompt and nonprompt multiplicity distributions. The nonprompt sample could determine the beauty jet response directly from data. The prompt sample would test the direct-channel mixtures with templates evaluated at the LHC energy and acceptance, including the remaining feed-down from $\chi_{cJ}$ and $\psi(2S)$ decays.

\medskip
{\small\noindent\textbf{Data availability.} The public experimental inputs are identified in Refs.~\cite{STAR:QWG2024jet,STAR:2020polarization,STAR:2018spectrum,PHENIX:2012psi,STAR:2013beauty,STAR:2019spectrum}. The simulated jet-response templates, numerical results, and analysis code are available from the corresponding author upon reasonable request.\par}
\smallskip
{\small\noindent\textbf{Acknowledgments.} We acknowledge support from Taiwan's National Science and Technology Council (NSTC), the Institute of Physics at Academia Sinica, and National Cheng Kung University. We thank STAR for its public preliminary results on charged-jet multiplicity in $J/\psi$ events at 200~GeV.\par}
\smallskip
{\small\noindent\textbf{AI Use Disclosure.} We used OpenAI ChatGPT and Codex (GPT-6 family for the final checks) to assist with analysis code, figures drawn from the numerical outputs, and the drafting and revision of the manuscript. The authors specified the jet definition and the benchmark comparisons. The reported values were checked against the numerical templates and fit outputs. We take full responsibility for the calculations, their interpretation, and the final text.\par}

\endgroup
\FloatBarrier
\begingroup\linespread{0.925}\footnotesize
\setlength{\bibsep}{0pt}

\endgroup
\end{document}